\documentclass[intlimits,twoside,a4paper]{article}

\usepackage[cp1251]{inputenc}
\usepackage{multirow}
\usepackage[eqsecnum]{cmpj3}

\usepackage{bm}

\issue{2023}{26}{2}{23601}
\doinumber{10.5488/CMP.26.23601}

\title[Random-anisotropy mixed-spin Ising on a triangular lattice] 
{Random-anisotropy mixed-spin Ising on a triangular lattice}	

\author[E. S. de Santana, A. S. de Arruda, M. Godoy] 
{E. S. de Santana\orcid{0000-0003-3058-2868}, A. S. de Arruda\orcid{0000-0002-7848-4514}, M. Godoy\orcid{0000-0001-9122-6061}\thanks{Corresponding author: \email{mgodoy@fisica.ufmt.br}}}	

\address{Instituto de F\'isica, Universidade Federal de Mato Grosso, 78060-900,
	Cuiab\'{a}, Mato Grosso, Brazil}


\date{Received June 21, 2022, in final form September 14, 2022}
\begin{document}
	\maketitle

\begin{abstract}
We have studied the mixed spin-1/2 and 1 Ising ferrimagnetic system
with a random anisotropy on a triangular lattice with three interpenetrating
sublattices $A$, $B$, and $C$. The spins on the sublattices are
represented by $\sigma_{A}$ (states $\pm1/2$), $\sigma_{B}$ (states
$\pm1/2$), and $S_{C}$ (states $\pm1$, $0$). We have performed Monte
Carlo simulations to obtain the phase diagram temperature $k_{\text{B}}T/\left|J\right|$
versus the strength of the random anisotropy $D/\left|J\right|$.
The phase boundary between two ferrimagnetic $FR_{1}$ and $FR_{2}$
phases at lower temperatures are always first-order for $p<0.25$
and second-order phase transition between the $FR_{1}$, $FR_{2}$
and the paramagnetic $P$ phases. On the other hand, for values of
$p\gtrapprox0.5$, the phase diagram presents only second-order phase
transition lines.
\printkeywords
\end{abstract}


\section{Introduction}

Many condensed matter researchers have studied models that describe
static magnetism in different materials. One of these models is the
well-known mixed-spin Ising model which can describe ferrimagnetic
materials \cite{1,2,3}. The interest in the study of ferrimagnetic
materials is due to their potential technological applications \cite{4}
which is a consequence of this material exhibiting a compensation
temperature ($T_{\text{comp}}$). This phenomenon occurs when the total magnetization
is zero at a temperature lower than the critical temperature ($T_{c}$).

The mixed-spin model can model the ferrimagnetic materials because
they are made up of repetitions of two different atoms with spins
of different magnitudes coupled antiferromagnetically each on a sublattice.
Your possible ground state can be that with all spins aligned antiferromagnetically
with a total magnetization greater than zero. The phase transition
from the ordered (ferrimagnetic) to the disordered (paramagnetic)
state occurs when the two sublattice magnetizations are zero (total
magnetization is zero) at a critical temperature $T_{c}$. On the
other hand, we have another interesting situation where the total
magnetization  can also be zero when the two sublattice magnetizations
are non-zero, i.e., the sum of both is zero. This point is known as
the compensation point ($T_{\text{comp}}$) and occurs at a temperature smaller
than the critical temperature ($T_{\text{comp}}<T_{c}$). Kaneyoshi et
al. \cite{5,6} and Plascak et al. \cite{7} performed
theoretical studies to understand the influence of the  anisotropy
on the magnetic properties and the compensation temperature in ferrimagnetic
materials.

There are a lot of studies carried out with different combinations
of spins, such as exact solutions~\cite{8,9,10,11} for the simplest
combination spin-1/2 and 1. Moreover, the mixed-spin Ising model has
been studied by different approaches such as mean-field approximation
\cite{12,13,14,15,16,17}, effective-field theory \cite{18,19,20,21},
re-normalization group~\cite{22}, numerical Monte Carlo simulations~\cite{23,24,25,26,27}.

An old controversy over the mixed-spin Ising model with  anisotropy
is related to the existence of a tricritical behaviour and a compensation
temperature. The origin of such controversy came from the works \cite{28,29,30,31}
carried out by using different approaches, where they indicated the
absence of such behaviours. These studies were later confirmed, only
for the two-dimensional case, by Selke and Oitmaa~\cite{32}, Godoy
et al. \cite{33}, and Leite et al. \cite{34,35}.
They performed Monte Carlo simulations in square and hexagonal lattices.
Therefore, these works concluded that the mixed-spin Ising model with
an anisotropy, the simplest version (spin-1/2 and 1) and in two dimensions,
does not exhibit tricritical behaviour. There is an exception, in a
very special case as shown by \u{Z}ukovi\u{c} and Bob\'{a}k \cite{35}, where
the two-dimensional lattice consists of three sublattices such that
a spin-1/2 is surrounded by the six spins-1 nearest neighbors. On
the other hand, Selke and Oitmaa \cite{32} showed that the model
in a cubic lattice exhibits such phenomena.

Another interesting controversy in this model is related to famous
magnetic frustration. Magnetic frustration is related to the fact
that the spins of some antiferromagnets are incapable of performing their
anti-parallel alignments simultaneously. This fact created great theoretical
and experimental challenges. More specifically, a triangular lattice
due to its geometry does not permit all interactions in an Ising type
system to be minimized simultaneously, which gives rise to the phenomenon
known as frustration. For example, consider three spins (nearest neighbors)
in a triangular lattice. Two of which are anti-parallel aligned,
satisfying their antiferromagnetic interactions ($J<0$), but the
third will never achieve such alignment simultaneously with the two
others. This phenomenon is known as magnetic frustration see~\cite{36}
and their references. Furthermore, the mixed-spin Ising model on a triangular
lattice, either in the ferrimagnetic or the antiferromagnetic case
produces frustration and it strongly changes the critical behaviour.
The interest in magnetic frustration in the lattice has grown and
it is now being considered in other models \cite{38,39}, such as
the Blume-Capel antiferromagnet \cite{40,41}. Another focus of interest
in triangular networks is the investigation of dynamic properties
of the kinetic Ising model with the use of Glauber-type stochastic
dynamics. In these studies, the temporal variation of the order parameter,
thermal behaviour of the total magnetization dynamics, the dynamics
of the phase diagram, and so on, are investigated \cite{42,43,44}.
In addition to these facts, the great interest in triangular lattices
is due to be useful for modelling some real materials, such as Ca$_{3}$Co$_{2}$O$_{6}$,
and CsCoX$_{3}$ (with X = Br or Cl) \cite{45,46}.

\u{Z}ukovi\u{c} and Bob\'{a}k \cite{46}, in another work more recently also performed Monte Carlo simulations in the mixed-spin (spin-1/2
and 1) Ising ferrimagnets model and focused only on its tricritical
behaviour induced by magnetic frustrations. These spins were distributed
on a triangular lattice (two-dimensional) with three sublattices $A$,
$B$, and $C$, where the spins can be arranged in two different ways
($\sigma_{A}$, $\sigma_{B}$, $S_{C}$) = ($1/2,1/2,1$). In this
work, they showed that the tricritical behaviour appears in a two-dimensional
ferrimagnets system, but in lattices whose topology has six nearest
neighbors. Thus, inspired by this work, we performed a Monte Carlo
simulation focusing our special attention only on the existence of
the first-order phase transition, but now the anisotropy is considered
to be random.

The paper is organized as follows: in section~\ref{seq2}, we  described
our system with a random  anisotropy on a triangular lattice and we
present some details concerning the simulation procedures. In section~\ref{seq3}, we  showed the results obtained. In section~\ref{seq4}, we presented
our conclusions.

\section{The model and simulations}\label{seq2}

\begin{figure}[h]
\centering{}\includegraphics[clip,scale=0.6]{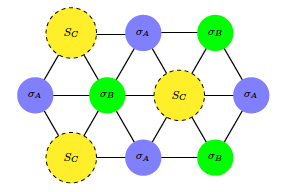} 
\caption{(Colour online) Schematic representation of the mixed spin-1/2 and
1 on a triangular lattice which consists of three interpenetrating
sublattices $A$, $B$, and $C$. The small and large circles denote
the spins-1/2 ($\sigma_{A}$, $\sigma_{B}$) and spin-1 ($S_{C}$)
sites, respectively. The full-straight lines represent the nearest
neighbor interaction $J<0$ and the dashed line denotes the existence
of the random  anisotropy.}
\label{fig1} 
\end{figure}

In order to study the behaviour of the thermodynamic quantities of
a mixed-spin Ising system on the triangular lattice with random  anisotropy,
we define the following Hamiltonian model,
\begin{align}
\mathcal{H}  =  -J\Bigl(\sum_{i\in A,j\in B}\sigma_{i}\sigma_{j}+\sum_{i\in A,k\in C}\sigma_{i}S_{k}+\sum_{j\in B,k\in C}\sigma_{j}S_{k}\Bigr)
  -\sum_{k\in C}D_{k}S_{k}^{2},\label{eq1}
\end{align}
where the spin variables assume the values $\sigma=\pm1/2$, $S=\pm1$
and 0, and the nearest-neighbor interaction is $J<0$. Here, there
are three different antiferromagnetic interaction types between the
nearest neighbor spins, such as $\sigma_{B}\leftrightarrow\sigma_{A}$,
$\sigma_{B}\leftrightarrow S_{C}$ and $\sigma_{A}\leftrightarrow S_{C}$
(see figure \ref{fig1}). They are distributed on three interpenetrating
sublattices $A$, $B$, and $C$. The random anisotropy $D_{k}$ in
the last term of equation~(\ref{eq1}) satisfies the following probability
distribution: 
\begin{equation}
P(D_{k})=p\delta(D_{k})+(1-p)\delta(D_{k}-D),\label{eq2}
\end{equation}
where the term $p\delta(D_{k})$ of the equation (\ref{eq2}) indicates
that a percentage $p$ of spins on the sublattice $C$ is free of
action of random anisotropy, whereas the term $(1-p)\delta(D_{k}-D)$
indicates a percentage ($1-p$) of spins on the sublattice $C$ under
the action of a random anisotropy.

In this work, the magnetic properties of the system are studied
using Monte Carlo simulations. In our simulations, we used linear
lattice sizes $L$ in the range of $L=48$ ($N=2304$ sites) to 105
($N=11025$ sites), and where $N=L^{2}$ is the number of lattice
sites. These lattices consist of three interpenetrating sublattices
with periodic boundary conditions. The initial states of the system
were prepared randomly and updated by the Metropolis algorithm \cite{47}.
We  used $10$ independent samples for any lattice size and $3.5\times10^{6}$
MCs (Monte Carlo steps) for the calculation of average values of thermodynamic
quantities of interest after discarding $1.5\times10^{6}$ MCs for
thermalization. 

We  calculated the sublattice magnetizations per site $m_{A}$,
$m_{B}$, and $m_{C}$ defined as 
\begin{equation}
m_{A(B)}=\frac{3}{L^{2}}\Bigl[\Bigl\langle \Bigl|\sum_{i\in A(B)}\sigma_{i}\Bigr|\Bigr\rangle \Bigr],\label{eq:3}
\end{equation}
and 
\begin{equation}
m_{C}=\frac{3}{L^{2}}\Bigl[\Bigl\langle \Bigl|\sum_{k\in C}S_{k}\Bigr|\Bigr\rangle \Bigr],\label{eq:4}
\end{equation}
where $\langle\cdots\rangle$ denotes the thermal average and $[\cdots]$
denotes the average over the sample of the system. Based on the ground-state
considerations (see below), for the identified ordered phases, we additionally
define the following order parameters for the entire system, which
take values between 0 in the fully disordered and 1 in the fully ordered
phase. Thus, we need to introduce two other order parameters (staggered
magnetization per site), $m_{s1}$ and $m_{s2}$, given by 
\begin{equation}
m_{s1}=\frac{1}{L^{2}}\Bigl[\Bigl\langle \Bigl|\sum_{k\in C}S_{k}-2\sum_{i\in A}\sigma_{i}-2\sum_{j\in B}\sigma_{j}\Bigr|\Bigr\rangle \Bigr],\label{eq:5}
\end{equation}
and 
\begin{equation}
m_{s2}=\frac{3}{L^{2}}\Bigl[\Bigl\langle \Bigl|\sum_{i\in A}\sigma_{i}-\sum_{j\in B}\sigma_{j}\Bigr|\Bigr\rangle \Bigr].\label{eq:6}
\end{equation}

Further, to find all the critical points, we  used the position
peak of the specific heat per site, given by 
\begin{equation}
c_{L}=\frac{[\langle E^{2}\rangle]-[\langle E\rangle]^{2}}{k_{\text{B}}T^{2}L^{2}},\label{eq:7}
\end{equation}
where $k_{\text{B}}$ is the Boltzmann constant, $E$ is the total energy
of the system and $L$ is the linear lattice size.

\section{Results and discussions}\label{seq3}

\begin{figure}
\centering
\includegraphics[clip,scale=0.35]{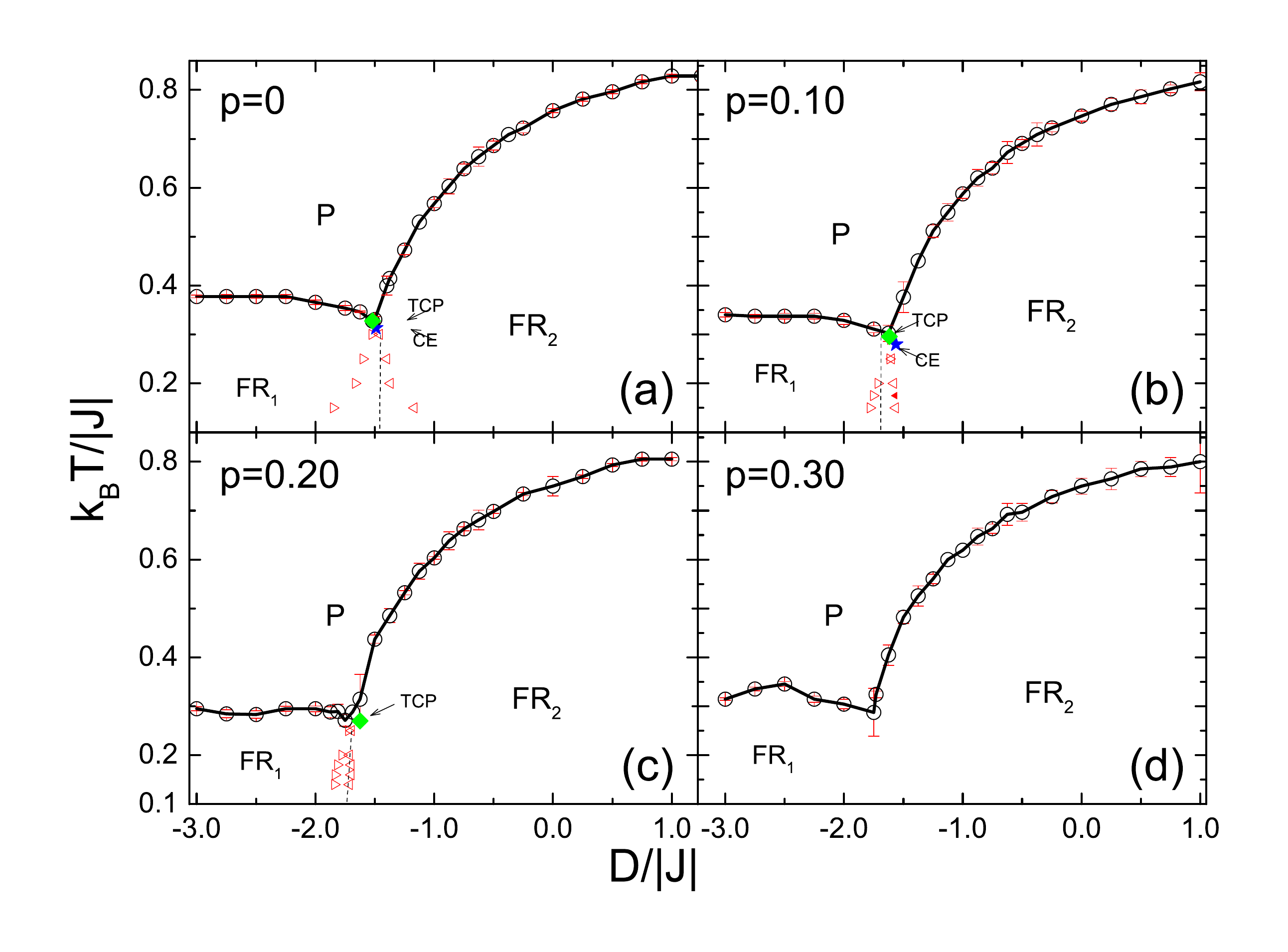}
\caption{(Colour online) Phase diagram in the $D/\left|J\right|$ versus $k_{\text{B}}T/\left|J\right|$
plane for different values of $p$: (a) $p=0$ (special case), (b)
$p=0.10$, (c) $p=0.20$, and (d) $p=0.30$. The empty circles represent
the phase transition temperatures $T_{c}$ between the paramagnetic
$P$ phase and the ferrimagnetic $FR_{1}$ and $FR_{2}$ phases calculated
by the specific heat peaks. The empty triangles denote the hysteresis
widths and the dotted lines represent an eye guide for the first-order
phase transition lines. The green-diamond and blue-star dots denote
the tricritical point (TCP) and the critical endpoint (CE), respectively.}
\label{fig2} 
\end{figure}

Firstly, we calculated all the possible ground-states for the entire
range of the  anisotropy parameter $D/\left|J\right|$. We  considered
the lattice of the system consisting of three interpenetrating sublattices
$A$, $B$, and $C$, as schematically defined in figure \ref{fig1}
and in the Hamiltonian, equation (\ref{eq1}). Focusing on a triangular
elementary unit cell consisting of the spins $\sigma_{A}$, $\sigma_{B}$,
$S_{C}$, one can obtain expressions for the reduced energies per
spin $E/(N\left|J\right|)$ of different spin arrangements as functions
of $D/\left|J\right|$. Then, the ground-states are determined as configurations
corresponding to the lowest energies for different values of $D/\left|J\right|$,
the ground-state configurations and the respective energies for different
ranges of the anisotropy parameter. Therefore, we  defined: (i)
the $FR_{1}$ phase with states ($\pm1/2,\mp1/2,0$) and energy $E/(N\left|J\right|)=-1/4$
for anisotropy in the range of $-\infty\leqslant D/\left|J\right|\leqslant-3/2$,
(ii) the $FR_{2}$ phase with states ($\pm1/2,\pm1/2,\mp1$) and energy
$\frac{E}{N\left|J\right|}=-3/4-\frac{D}{3\left|J\right|}$ for anisotropy
in the range of $-3/2\leqslant D/\left|J\right|\leqslant+\infty$. On the other
hand, for $k_{\text{B}}T/\left|J\right|\neq0$ besides the ordered ferrimagnetic
$FR_{1}$ and $FR_{2}$ phases (similar at $k_{\text{B}}T/\left|J\right|=0$),
we also have another phase, i.e., the paramagnetic $P$ phase. We can note
that in the $FR_{1}$ phase, the zero means nonmagnetic states $S_{k}=0$
of spins on sublattice $C$. The critical value of the anisotropy
parameter separating the respective $FR_{1}$ and $FR_{2}$ phases
is given by $D_{c}/\left|J\right|=-3/2$.

\begin{figure}
\centering{}\includegraphics[clip,scale=0.3]{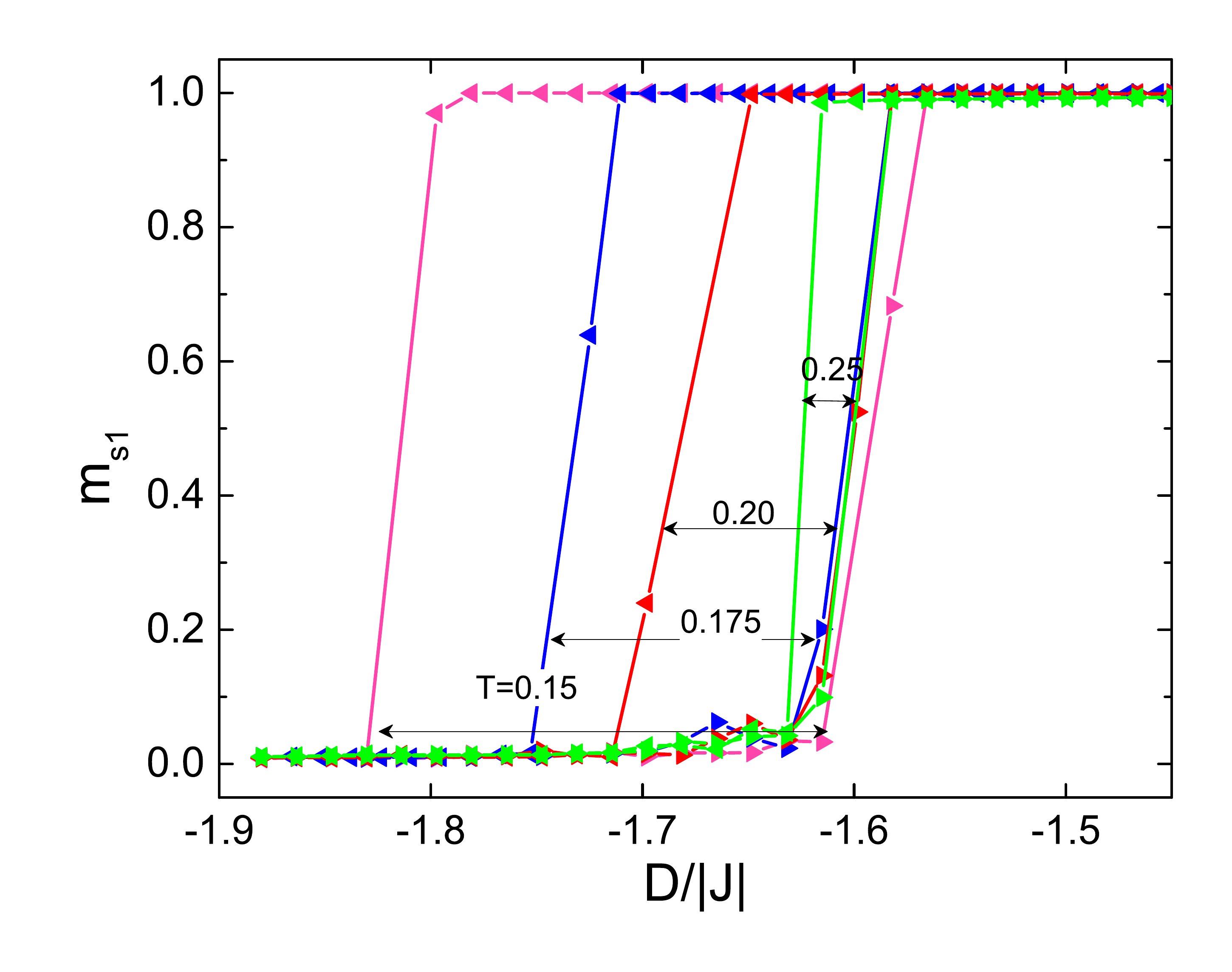}
\caption{(Colour online) The hysteresis loops of the system, staggered magnetization
$m_{s1}$ as a function of the strength of the random  anisotropy
$D/\left|J\right|$. The symbols $\blacktriangleright$ and $\blacktriangleleft$
indicate that $m_{S1}$  increases and decreases with strength
of $D/\left|J\right|$, respectively. These loops were obtained for
different values of temperatures $k_{\text{B}}T/\left|J\right|$ as indicated
in the figure. Here, we  used $L=66$ ($N=4356$ sites) and $p=0.10$.
The error bars are within the symbol size.}
\label{fig3} 
\end{figure}

Now, let us turn our attention to  figure \ref{fig2}, where we 
presented the phase diagram in the $D/\left|J\right|$ versus $k_{\text{B}}T/\left|J\right|$
plane for some select values of $p\leqslant0.3$, for instance, for $p=0$
(special case, see figure~\ref{fig2}(a)), $p=0.10$ (figure \ref{fig2}(b)),
$p=0.20$ (figure \ref{fig2}(c)), and $p=0.30$ (figure \ref{fig2}(d)).
The results for the special case $p=0$ in figure \ref{fig2}(a) were
obtained in a  way similar to the ones obtained by \u{Z}ukovi\u{c} and Bob\'{a}k~\cite{46}
and we  reproduced the results here. This case is related to the
situation that all the spins on the sublattice $C$ are under the
influence of  anisotropy $D/\left|J\right|$. In this case, the topology
of the phase diagram presents three different phases: the paramagnetic
$P$ phase, the ferrimagnetic $FR_{1}$ phase and another ordered
phase also ferrimagnetic $FR_{2}$ phase. The phase transitions between
the $P-FR_{1}$ and $P-FR_{2}$ phases are continuous phase transitions
(second-order phase transition). The empty circles represent the phase
transition temperatures $k_{\text{B}}T_{c}/\left|J\right|$ between the phases
and are estimated by the specific heat peak $c$, equation (\ref{eq:7}).
We  found $k_{\text{B}}T_{c}/\left|J\right|=0.757\pm0.001$ for $D/\left|J\right|=0$
and it is approximately equal to the results obtained by \u{Z}ukovi\u{c}. We 
can also observe a first-order transition line between the ordered $FR_{1}-FR_{2}$
phases at low temperature. We  used the same technique that was
used by \u{Z}ukovi\u{c} which consists of plotting the order parameter (in
our case, the staggered magnetization $m_{s1}$, equation (\ref{eq:5})
increasing and decreasing as a function of $D/\left|J\right|$. Through
these plots, we can observe its discontinuous character by the appearance
of hysteresis loops, and when the width of the hysteresis loops increases
for lower temperatures. Similar behaviour is demonstrated in figure \ref{fig3}
for the magnetization $m_{s1}$ and $p=0.10$, where such behaviour
signals a first-order phase transition. These hysteresis loops persist
for higher temperatures, but this behaviour disappears when the temperature
increases and is close to the critical endpoint (CE) and the tricritical
point (TCP). The highest temperatures at which we still could observe
some signs of first-order phase transitions  after this behaviour
disappear. Thereby, we can estimate the coordinate of the CE and the
TCP points. We  found for the coordinate of the CE ($D_{\text{CE}}/\left|J\right|=-1.618$,
$k_{\text{B}}T_{\text{CE}}/\left|J\right|=0.296$) and the TCP ($D_{\text{TCP}}/\left|J\right|=-1.520$,
$k_{\text{B}}T_{\text{TCP}}/\left|J\right|=0.327$) points for the case $p=0$.

When we decrease the number of spins on the sublattice $C$ under
the action of the anisotropy $D/\left|J\right|$, that is, the value
of $p$ becomes greater (see figure \ref{fig2}(b) with $p=0.10$ (10\%)),
this anisotropy makes the critical temperature to decrease. Therefore,
it induces the first-order transition line for smaller values of $D/\left|J\right|$
and temperature in the range of $0.3\geqslant k_{\text{B}}T/\left|J\right|\geqslant0.2$.
Here, we can see that the transition again becomes second-order for
large values of $D/\left|J\right|$. In this case, the CE and the
TPC points are located only in regions of lower temperatures where
the coordinates are ($D_{\text{CE}}/\left|J\right|=-1.563$, $k_{\text{B}}T_{\text{CE}}/\left|J\right|=0.280$)
and ($D_{\text{TCP}}/\left|J\right|=-1.625$, $k_{\text{B}}T_{\text{TCP}}/\left|J\right|=0.302$),
respectively. Moreover, in figure \ref{fig2}(c) we can observe a decrease
in the region of hysteresis loops where there is the first-order phase transition.
For this case with $p=0.20$, the coordinate of the TCP is ($k_{\text{B}}D_{\text{TCP}}/\left|J\right|=-1.75$,
$k_{\text{B}}T_{\text{TCP}}/\left|J\right|=0.27$). Here, due to the difficulties
in the simulations, we did not obtain the coordinates of the CE point.
On the other hand, for the case $p=0.30$ shown in figure \ref{fig2}(d),
we did not find a first-order phase transition between the $FR_{1}$
and $FR_{1}$ phases using the hysteresis loop technique. We 
found a critical parameter $p_{c}=0.25$, where the first-order phase
transition line $FR_{1}-FR_{2}$ disappears.

\begin{figure}
\centering{}\includegraphics[viewport=0bp 0bp 414bp 400bp,scale=0.53]{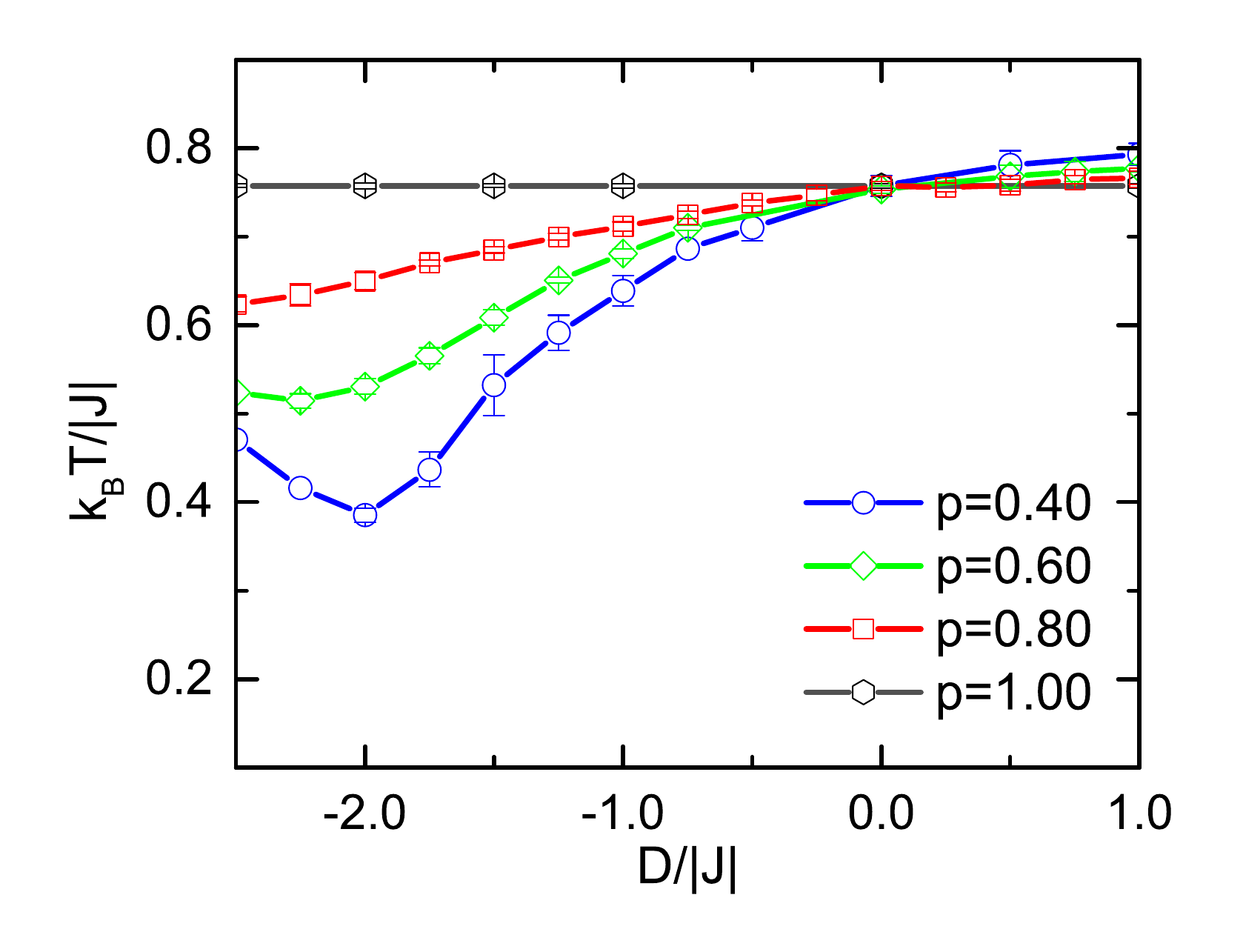}
\caption{(Colour online) The phase diagram in the $D/\left|J\right|$ versus
$k_{\text{B}}T/\left|J\right|$ plane for values of $0.4\protect\leqslant p\protect\leqslant1.0$,
as indicated in the figure. The full lines represent the phase transition
lines between the paramagnetic $P$ and the ferrimagnetic $FR_{2}$
phases.}
\label{fig4} 
\end{figure}

A behaviour hysteresis loop characteristic is shown in figure \ref{fig3},
where we exhibit the magnetization $m_{s1}$ as a function of an
increase and decrease of the strength of the random  anisotropy parameter
$D/\left|J\right|$ for different values of temperatures such as $k_{\text{B}}T/\left|J\right|=0.15$,
0.175, 0.20, 0.25. When we increase the temperature, the area of the
hysteresis loop decreases until the disappearance of $k_{\text{B}}T/\left|J\right|\geqslant0.30$.

For other values of $p$ in the range of $0.40\leqslant p\leqslant1.00$, we
also  illustrated in figure \ref{fig4} the phase diagram in the
$D/\left|J\right|$ versus $k_{\text{B}}T/\left|J\right|$ plane. The empty
symbols represent the phase transition temperatures $k_{\text{B}}T_{c}/\left|J\right|$
between the paramagnetic $P$ and the ferrimagnetic $FR_{2}$ phases
and they are all second-order phase transitions. All the values were
estimated by the specific heat peak, equation (\ref{eq:7}). Another important
point we want to emphasize is that for $p=1.00$ where $k_{\text{B}}T_{c}/\left|J\right|$
is independent of the anisotropy $D/\left|J\right|$.

\begin{figure}
\centering{\includegraphics[clip,scale=0.3]{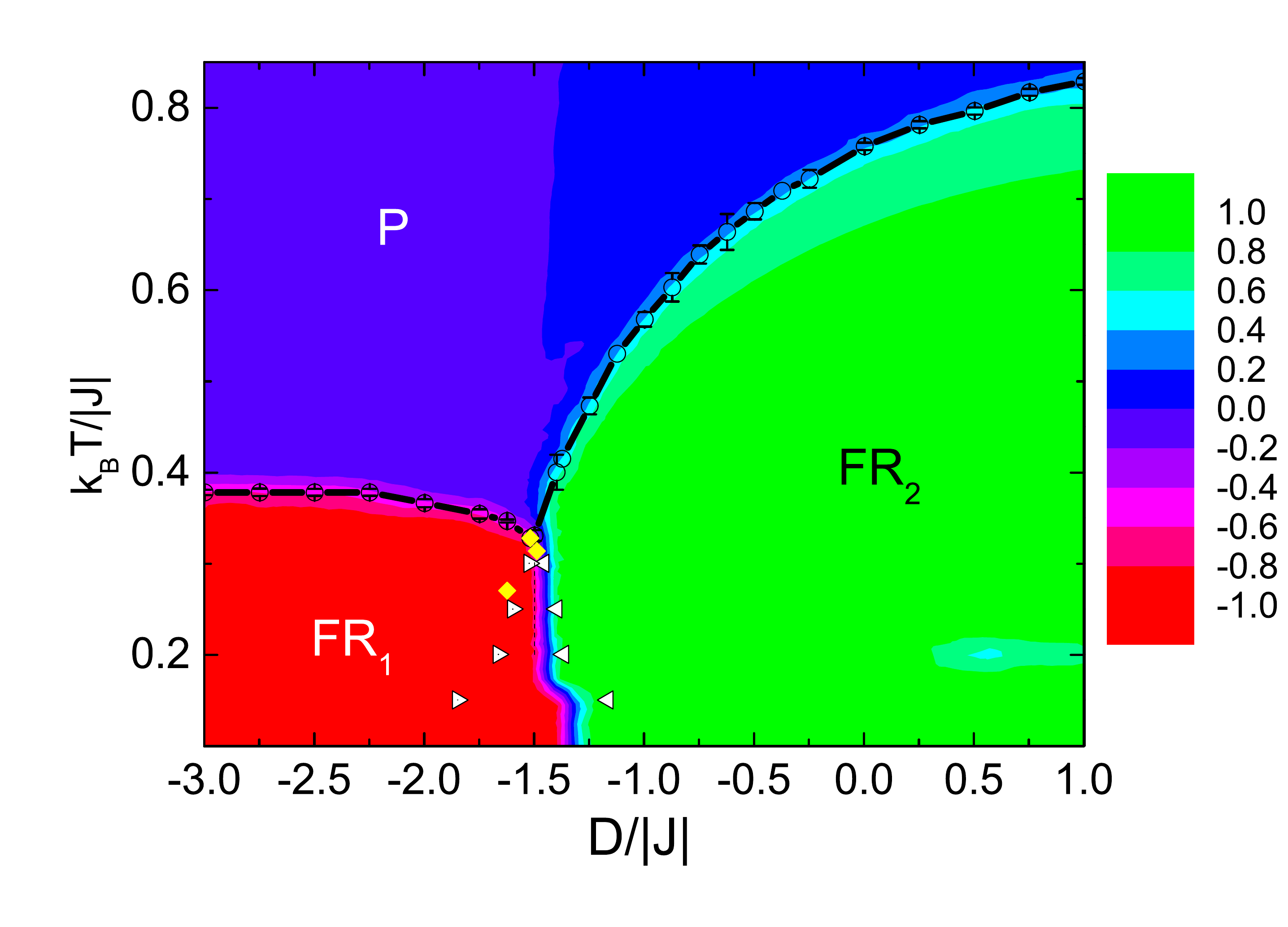}} 
\caption{(Colour online) Phase diagram with different background colours in the
$k_{\text{B}}T/\left|J\right|$ versus $D/\left|J\right|$ plane for the
case $p=0$. The red colour  is the ferrimagnetic $FR_{1}$ phase, green
is the ferrimagnetic $FR_{2}$ phase, and blue is the paramagnetic
$P$ phase.}
\label{fig5} 
\end{figure}

Due to the great difficulty in simulating at low temperatures and
to have a complete idea of what happens with the different phase regions
in the phase diagram when we change $p$, we decided to construct
a phase diagram with background colours. Thereby, we can visualize
the effects of the strength of the  anisotropy dilution in the phase
topology of the phase diagrams. To this end, we  constructed this
diagram using the following definition for the parameter, $M_{d}=m_{s1}-m_{s2}$.
This new parameter has a range of ($-1.0$) to (1.0) where we can assign
a colour spectrum as can be shown in figure \ref{fig5}. When $m_{s1}\rightarrow1.0$
and $m_{s2}\rightarrow0$ so $M_{d}\rightarrow1.0$, on the other
hand, when $m_{s1}\rightarrow0$ and $m_{s2}\rightarrow1.0$ so $M_{d}\rightarrow-1.0$.
Using this definition, we have presented in figure \ref{fig5} an example
of the phase diagram with different background colours in the $D/\left|J\right|$
versus $k_{\text{B}}T/\left|J\right|$ plane for the case $p=0$. The red colour
 denotes the ferrimagnetic $FR_{1}$ phase ($M_{d}\approx-1.0$),
green is the ferrimagnetic $FR_{2}$ phase ($M_{d}\approx1.0$) and
blue is the paramagnetic $P$ phase ($M_{d}\approx0$). As we can
see, the dominant colours correspond to the areas of the phases as seen
in figure \ref{fig2}(a) (case $p=0$) and which correspond to the ferrimagnetic
$FR_{2}$ phase (green), the ferrimagnetic $FR_{1}$ phase (red) and
the paramagnetic $P$ phase (blue). To make it clearer, we also plot
the data of figure~\ref{fig2}(a) in figure \ref{fig5}.

\begin{figure}
\begin{centering}
\includegraphics[viewport=0bp 0bp 770bp 1100bp,scale=0.34]{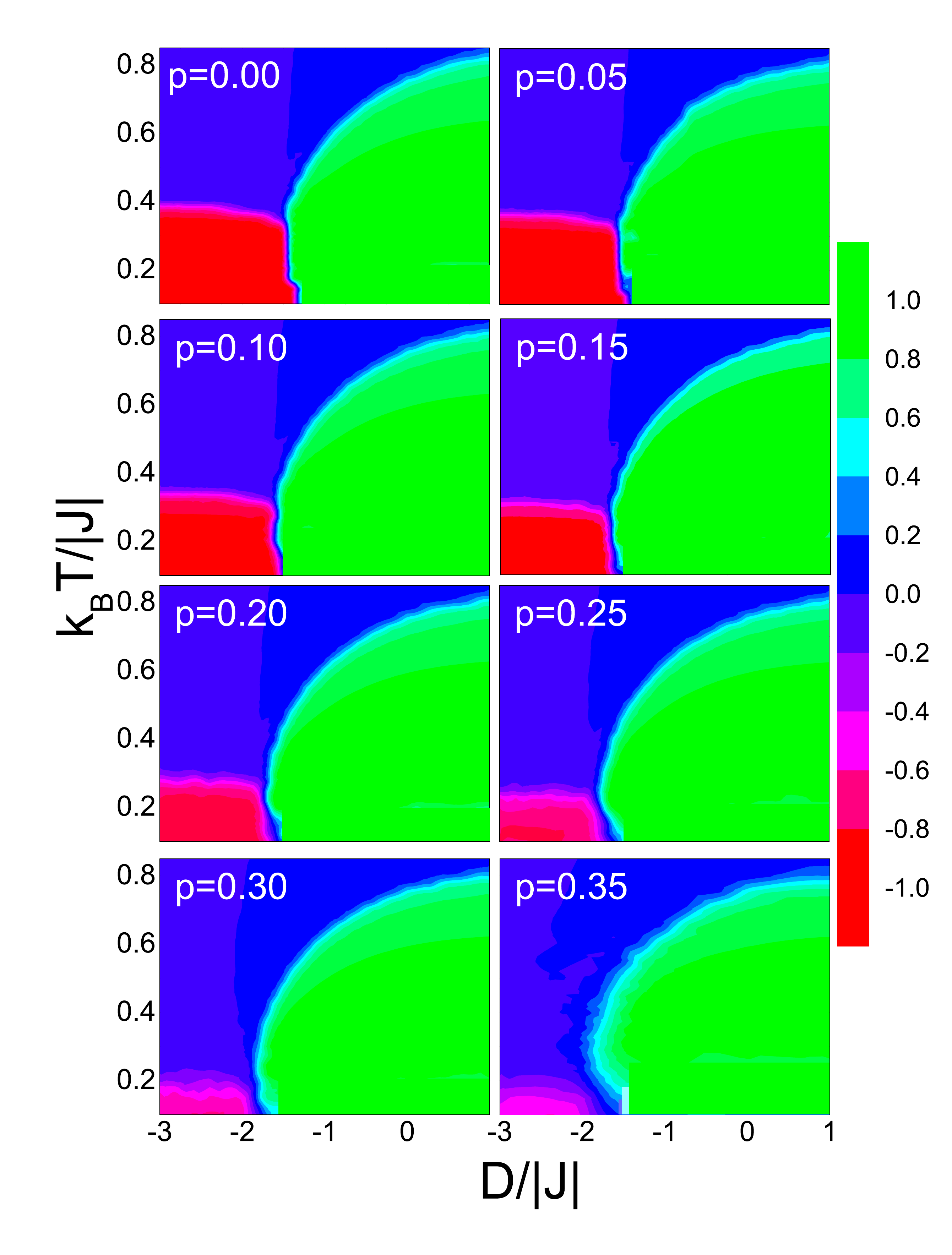} 
\par\end{centering}
\centering{}\caption{(Colour online) Phase diagram with different background colours in the
$k_{\text{B}}T/\left|J\right|$ versus $D/\left|J\right|$ for values of
$0\protect\leqslant p\protect\leqslant0.35$, as indicated in the figures. The
red colour is the ferrimagnetic $FR_{1}$ phase, green is the ferrimagnetic
$FR_{2}$ phase, and blue is the paramagnetic $P$ phase.}
\label{fig6} 
\end{figure}

\begin{figure}
\centering{}\includegraphics[viewport=0bp 0bp 776bp 789bp,clip,scale=0.32]{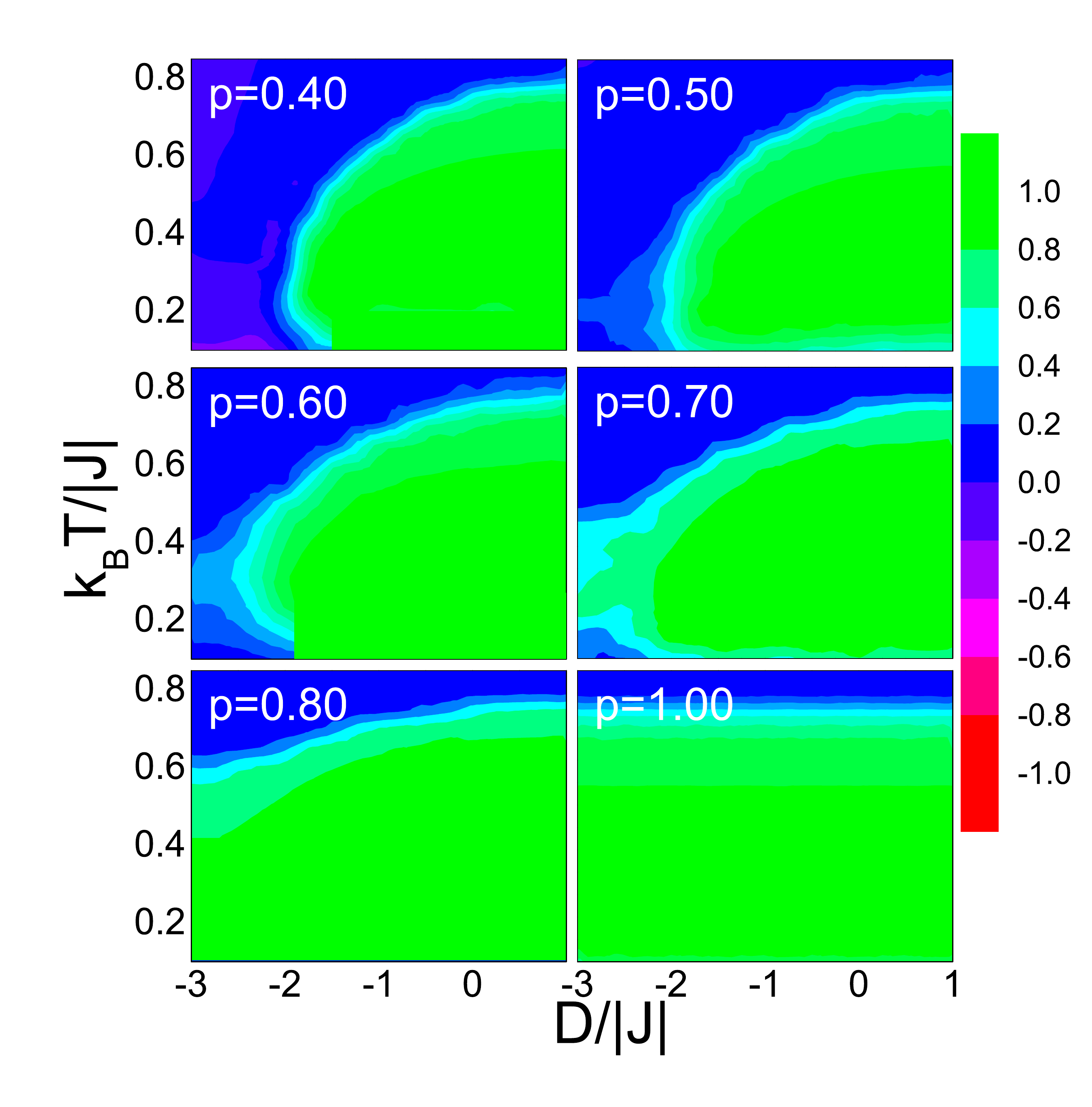}
\caption{(Colour online) Phase diagram with different background colours in the
$k_{\text{B}}T/\left|J\right|$ versus $D/\left|J\right|$ for values of
$0.4\protect\leqslant p\protect\leqslant1.0$, as indicated in the figures.
The colour green is the ferrimagnetic $FR_{2}$ phase and blue is the
paramagnetic $P$ phase.}
\label{fig7} 
\end{figure}

Thus, in figure \ref{fig6} we  exhibited the phase diagram with
background colours in the $k_{\text{B}}T/\left|J\right|$ versus $D/\left|J\right|$
plane for several values of $p$ in the range of $0\leqslant p\leqslant0.35$.
Here, the colours continue to be red for the ferrimagnetic $FR_{1}$
phase, green for the ferrimagnetic $FR_{2}$ phase, and blue for the
paramagnetic $P$ phase. For the case $p=0$, we have the same topology
as presented in figure \ref{fig5} and obtained by \u{Z}ukovi\u{c} \cite{46}.
On the other hand, when $p$ increases ($p\geqslant0.05$) we can observe
that the $FR_{1}$ phase decreases more and more (see $p=0.30$ and
0.35) and then for $p\gtrapprox0.4$ (see in figure \ref{fig7}) the
phase diagram presents only two $P$ and $FR_{2}$ dominant phases.
This case means that we have approximately $40\%$ or more spins under
the influence of random anisotropy $D/\left|J\right|$.

Finally, in figure \ref{fig7}, we  displayed the phase diagram
which is a continuation of figure \ref{fig6}, but now for values of
$p$ in the range of $0.4\leqslant p\leqslant1.0$. In this case, we still have
only two phases, the $P$ and $FR_{2}$ phases. When $p$ increases,
the $FR_{2}$ phase grows for $D/\left|J\right|\rightarrow-\infty$.
This growth starts at $p\gtrapprox0.4$ and goes on increasing until
it occupies all the region for $k_{\text{B}}T/\left|J\right|=0.757$ in the case
$p=1.0$. All of these results corroborate the results shown in figure
\ref{fig4}, i.e. We can see that the phase boundary that separates
the $P$ from the $FR_{2}$ phase region coincides with the phase
separation line in figure \ref{fig4} for $p=1.0$.

\section{Conclusions}\label{seq4}

In conclusion, we  used Monte Carlo simulations to study the phase
diagram of the mixed spin-1/2 and 1 Ising ferrimagnetic system with
a random anisotropy on a triangular lattice with three interpenetrating
sublattices $A$, $B$, and $C$. The spins on the sublattices are
represented by $\sigma_{A}$, $\sigma_{B}$, $S_{C}$ with states
$\pm1/2$, $\pm1/2$, $\pm1$ and 0, respectively. We  obtained
the phase diagram at the temperature $k_{\text{B}}T/\left|J\right|$ versus
the strength of the random anisotropy $D/\left|J\right|$ plane, where
the anisotropy is randomly distributed on the sublattice $C$ (with
spins $S_{C}$) according to the bimodal probability distribution
$P(D_{k}/\left|J\right|)$. Therefore, we can conclude that the phase
boundary in the phase diagram presents a topology that depends on
the parameter $p$. We  found a first-order transition line between
the two ferrimagnetic $FR_{1}$ and $FR_{2}$ phases at lower temperatures
for $p<0.25$ together with a second-order phase transition between
the $FR_{1}-P$ and $FR_{2}-P$ phases. On the other hand, for $p\gtrapprox0.4$,
i.e., above $40\%$ of the sites on the sublattice $C$ are free of
$D/\left|J\right|$. The phase diagram presents only second-order
phase transition lines between ordered and paramagnetic $P$ phase
showing that the system no longer exhibits the tricritical behaviour.

\section{Acknowledgements}

The authors acknowledge financial support from the Brazilian agencies
CNPq and CAPES.

\ukrainianpart

\title[Випадкова анізотропія у системі зі змішаними спінами Ізинга на трикутній ґратці] 
{Випадкова анізотропія у системі зі змішаними спінами Ізинга на трикутній ґратці}	
\author[Е. С. де Сантана, А. С. де Арруда, М. Годой] 
{Е. С. де Сантана, А. С. де Арруда, М. Годой}	
\address{Інститут фізики, Федеральний університет Мату-Гросу, 78060-900, Куяба, Мату-Гросу, Бразилія}

\makeukrtitle

\begin{abstract}
	Досліджено феррімагнітну систему суміші Ізингових спінів-1/2 та 1 із випадковою анізотропією на три\-кут\-ній ґратці з трьома
	взаємопроникаючими підґратками $A$, $B$ та $C$. Спіни на підґратках задаються як $\sigma_{A}$ (стани $\pm1/2$), $\sigma_{B}$ (стани	$\pm1/2$), та $S_{C}$ (стани $\pm1$, 0). Проведено моделювання Монте-Карло для отримання фазової діаграми ``температура $k_{\text{B}}T/\left|J\right|$ -- випадкова анізотропія $D/\left|J\right|$''. Межа розділу між двома феррімагнітними фазами $FR_{1}$ та $FR_{2}$ за нижчих температур завжди відповідає фазовому переходу першого (для $p<0,25$) та другого роду (між $FR_{1}$, $ FR_{2}$ та парамагнітною $P$ фазами). З іншого боку, для значень $p\gtrapprox0.5$ фазова діаграма містить лише лінії фазових переходів другого роду.
	\keywords випадкова анізотропія, системи Ізинга зі змішаними спінами, моделювання Монте-Карло
\end{abstract}

\begin{thebibliography}{99}
\bibitem[1]{1}  Mallah T.,  Thiebaut S.,  Verdaguer M.,   Veillet P.,
Science, 1993, \textbf{262}, 1554, \doi{10.1126/science.262.5139.1554}.

\bibitem[2]{2}  Okawa H.,  Matsumoto N.,  Tamaki H.,   Ohba M., Mol. Cryst. Liq. Cryst. Lett., 1993, \textbf{233}, 257, \\\doi{10.1080/10587259308054965}.

\bibitem[3]{3} Mathoniere  C.,  Nutall C. J.,  Carling S. G.,  
Day P., Inorg. Chem., 1996, \textbf{35}, 1201, \doi{10.1021/ic950703v}.

\bibitem[4]{4}  Kahn O., Molecular Magnetism, VCH-Verlag, Weinheim, New York, 1993.

\bibitem[5]{5}  Kaneyoshi T.,   Nakamura Y., J. Phys.: Condens. Matter, 1998,
\textbf{10}, 3003, \doi{10.1088/0953-8984/10/13/017}.

\bibitem[6]{6}  Kaneyoshi T.,  Nakamura Y.,   Shin S., J. Phys.:Condens. Matter, 1998, \textbf{10}, 7025, \\\doi{10.1088/0953-8984/10/31/018}.

\bibitem[7]{7}  Verona de Resende H. F.,  S\'{a} Barreto F. C.,   Plascak J. A., Physica A, 1988, \textbf{149}, 606, \\\doi{10.1016/0378-4371(88)90121-5}.

\bibitem[8]{8}  Goncalves L. L., Phys. Scr., 1985, \textbf{32}, 248, \doi{10.1088/0031-8949/32/3/012}.

\bibitem[9]{9}  Lipowski A.,   Horiguchi T., J. Phys. A: Math. Gen., 1995, \textbf{28}, L261,  \doi{10.1088/0305-4470/28/9/003}.

\bibitem[10]{10}  Ja\u{s}\u{c}ur M., Physica A, 1998, \textbf{252}, 217, \doi{10.1016/S0378-4371(97)00584-0}.

\bibitem[11]{11}  Dakhama A., Physica A, 1998, \textbf{252}, 225, \doi{10.1016/S0378-4371(97)00583-9}.

\bibitem[12]{12}  Souza I. J.,  de Arruda P. H. Z.,  Godoy M.,  Craco L.,
 de Arruda  A. S., Physica A, 2016, \textbf{444}, \\589, \doi{10.1016/j.physa.2015.10.089}.

\bibitem[13]{13}  Da Cruz Filho J. S.,  Tunes T. M., Godoy  M.,
 de Arruda A. S., Physica A, 2016, \textbf{450}, 180, \\\doi{10.1016/j.physa.2015.12.096}.

\bibitem[14]{14}  Da Cruz Filho J. S.,  Godoy M.,   de Arruda A. S.,
Physica A, 2013, \textbf{392}, 6247, \doi{10.1016/j.physa.2015.12.096}.

\bibitem[15]{15}  Reyes J. A.,  de La Espriella N.,   Buend\'ia G. M.,
Phys. Status Solidi B, 2015, \textbf{10}, 252, \doi{10.1002/pssb.201552110}.

\bibitem[16]{16} De La Espriella  N.,  Mercado C. A.,   Madera J. C.,
J. Magn. Magn. Mater., 2016, \textbf{401}, 22, \\\doi{10.1016/j.jmmm.2015.09.083}.

\bibitem[17]{17}  Abubrig O. F.,  Horv\'ath D.,   Bob\'{a}k A.,  Ja\u{s}\u{c}ur M.,
Physica A, 2001, \textbf{296}, 437, \\\doi{10.1016/S0378-4371(01)00176-5}.

\bibitem[18]{18}  Kaneyoshi T., Physica A, 1988, \textbf{153}, 556,
\doi{10.1016/0378-4371(88)90240-3}.

\bibitem[19]{19}  Kaneyoshi T., J. Magn. Magn. Mater., 1990, \textbf{92}, 59, \doi{10.1016/0304-8853(90)90679-K}.

\bibitem[20]{20}  Benyoussef A.,  El Kenz A.,   Kaneyoshi T., J. Magn. Magn. Mater., 1994, \textbf{131}, 173, \\\doi{10.1016/0304-8853(94)90025-6}.

\bibitem[21]{21}  Benyoussef A.,  El Kenz A.,   Kaneyoshi T., J. Magn. Magn. Mater., 1994, \textbf{131}, 179, \\\doi{10.1016/0304-8853(94)90026-4}.

\bibitem[22]{22}  Quadros S. G. A.,   Salinas S. R., Physica A, 1994, \textbf{206},
479, \doi{10.1016/0378-4371(94)90319-0}.

\bibitem[23]{23}  Zhang G. M.,   Yang Ch. Z., Phys. Rev. B, 1993, \textbf{48},
9452, \doi{10.1103/PhysRevB.48.9452}.

\bibitem[24]{24}  Buendia G. M.,   Liendo J. A., J. Phys.: Condens.
Matter, 1997, \textbf{9}, 5439,  \doi{10.1088/0953-8984/9/25/011}.

\bibitem[25]{25}  Godoy M.,   Figueiredo W., Phys. Rev. E, 2000, \textbf{61},
218, \doi{10.1103/PhysRevE.61.218}.

\bibitem[26]{26}  Pereira J. R. V.,  Tunes T. M.,  de Arruda A. S., 
 Godoy M., Physica A, 2018, \textbf{500}, 265, \\\doi{10.1016/j.physa.2018.02.085}.

\bibitem[27]{27}  Da Silva D. C.,  de Arruda A. S.,  Godoy  M., Int.
J. Mod. Phys. C, 2020, \textbf{31}, No.~9, 2050124, \\\doi{10.1142/S0129183120501247}.

\bibitem[28]{28}  Kaneyoshi T.,   Chen J. C., J. Magn. Magn. Mater., 1991,
\textbf{98}, 201, \doi{10.1016/0304-8853(91)90444-F}.

\bibitem[29]{29}  Kaneyoshi T., J. Phys. Soc. Jpn., 1987, \textbf{56}, 2675,
\doi{10.1143/JPSJ.56.2675}.

\bibitem[30]{30}  Bob\'{a}k A.,   Jur\u{c}i\u{s} M., Physica A, 1997, \textbf{240},
647, \doi{10.1016/S0378-4371(97)00044-7}.

\bibitem[31]{31}  Oitmaa J.,  Enting I. G., J. Phys.: Condens. Matter, 2006,
\textbf{18}, 10931, \doi{10.1088/0953-8984/18/48/020}.

\bibitem[32]{32}  Selke W.,   Oitmaa J., J. Phys.: Condens. Matter, 2010,
\textbf{22}, 076004, \doi{10.1088/0953-8984/22/7/076004}.

\bibitem[33]{33}  Figueiredo W.,  Godoy M.,   Leite V. S., Braz. J. Phys., 2004, \textbf{34}, No.~2A, \doi{10.1590/S0103-97332004000300010}.

\bibitem[34]{34}  Leite V. S.,  Godoy M.,   Figueiredo W., Phys.
Rev. B, 2005, \textbf{71}, 094427, \doi{10.1103/PhysRevB.71.094427}.

\bibitem[35]{35}  \u{Z}ukovi\u{c} M.,   Bob\'{a}k A., Physica A, 2015, \textbf{436},
509, \doi{10.1016/j.physa.2015.05.077}.

\bibitem[36]{36}  Ertas M.,  Kocakaplan Y.,  Kantar E., J. Magn. Magn. Mater., 2015, \textbf{386}, No.~1--7, \\\doi{10.1016/j.jmmm.2015.03.058}.

\bibitem[37]{37}  Stre\v{c}ka J.,  Rebic M.,  Rojas O., de Souza  S. M.,
J. Magn. Magn. Mater., 2019, \textbf{469}, 655, \\\doi{10.1016/j.physleta.2019.05.017}.

\bibitem[38]{38}  Zad H. A.,   Ananikian N., J. Phys.: Condens. Matter, 2018,
\textbf{30}, 165403, \doi{10.1088/1361-648X/ab3136}.

\bibitem[39]{39}  \u{Z}ukovi\u{c} M.,   Bob\'{a}k A., Phys. Rev. E, 2013, \textbf{87},
032121, \doi{10.1103/PhysRevE.87.032121}.

\bibitem[40]{40}  Theodorakis P. E.,   Fytas N. G., Phys. Rev. E, 2012,
\textbf{86}, 011140,  \doi{10.1103/PhysRevE.86.011140}.

\bibitem[41]{41}  Kantar E.,  Ertas M., Phase Transitions, 2018, \textbf{91},
No.~4, 370--381, \doi{10.1080/01411594.2017.1402897}.

\bibitem[42]{42}  Ertas M.,  Kantar E., J. Supercond. Novel Magn., 2015, \textbf{28}, 3037--3044, \doi{10.1007/s10948-015-3134-2}.

\bibitem[43]{43}  Ertas M.,  Kantar E.,  Kocakaplan Y., Keskin M., Physica A, 2016, \textbf{444}, 732, \doi{10.1016/j.physa.2015.10.069}.

\bibitem[44]{44} Kudasov Y. B., Phys. Rev. Lett., 2006, \textbf{96}, 027212, \doi{10.1103/PhysRevLett.96.027212}.

\bibitem[45]{45}  Soto R.,  Martinez G.,  Baibich M. N.,  Florez J. M.,
  Vargas P., Phys. Rev. B, 2009, \textbf{79}, 184422, \\\doi{10.1103/PhysRevB.79.184422}.

\bibitem[46]{46}  \u{Z}ukovi\u{c} M.,   Bob\'{a}k A., Phys. Rev. E, 2015, \textbf{91},
052138, \doi{10.1103/PhysRevE.91.052138}.

\bibitem[47]{47}  Metropolis N.,  Rosenbluth A.,  Rosenbluth M., 
Teller A.,   Teller E., J. Chem. Phys., 1953, \textbf{21}, 1087, \doi{10.1063/1.1699114}.
\end{thebibliography}
\end{document}